\documentclass[prb,twocolumn]{revtex4-1}
\usepackage[table, dvipsnames]{xcolor}
\usepackage{graphicx}
\usepackage{amssymb}
\usepackage{dcolumn}
\usepackage{float}
\usepackage{bm}
\usepackage{multirow}
\usepackage{booktabs}
\usepackage[T1]{fontenc}
\usepackage[utf8]{inputenc}
\usepackage{lmodern}
\usepackage{color}
\usepackage{makecell}
\usepackage{tikz}
\usepackage{pgfplots}
\usepackage{mathtools}
\usepackage{amsmath} 
\usepackage{amssymb}
\usepackage{gensymb}

\usepackage[colorlinks=true,linkcolor=red]{hyperref}
\newcommand\scalemath[2]{\scalebox{#1}{\mbox{\ensuremath{\displaystyle #2}}}}

\newcommand{\mathleft}{\@fleqntrue\@mathmargin0pt}

\newcolumntype{M}[1]{>{\centering\arraybackslash}m{#1}}
\renewcommand{\Re}{\mathrm{Re}\,}
\renewcommand{\Im}{\mathrm{Im}\,}

\newcommand{\tens}{%
	\mathbin{\mathop{\otimes}}%
}
\DeclareMathAlphabet{\bi}{OML}{cmm}{b}{it}

\def\be{\begin{equation}}
\def\ee{\end{equation}}
\def\bearr{\begin{eqnarray}}
\def\eearr{\end{eqnarray}}

\begin{document}
\title{Quantum anomalous Hall phase and effective in-plane Lande-\textit{g} factor in an inverted InAs/GaSb quantum well}
\bigskip
\author{Sushmita Saha}
\email{phd2101251005@iiti.ac.in}
\author{Alestin Mawrie}
\email{amawrie@iiti.ac.in}
\normalsize
\affiliation{Department of Physics, Indian Institute of Technology Indore, Simrol, Indore-453552, India}
\date{\today}

	\begin{abstract}
		The inverted band structure discovered in InAs/GaSb quantum well (QW) is found to host the topological quantum spin Hall (QSH) states. A QSH insulator hosts counterpropagating spin-polarized edge states that are protected by the time-reversal symmetry. The latest experiment reported a robust quantized Hall conductance arising from these QSH states that persists in an in-plane magnetic field as strong as $12$ Tesla. Based on the result of this experiment, we present here a precise calculation of the effective in-plane Lande-\textit{g} factor. We based our calculations on the tight-binding Hamiltonian projected on a square lattice that reproduces a slightly modified Bernevig-Hughes-Zhang (BHZ) Hamiltonian. We also study the topological phase transitions \textit{w.r.t.} a magnetic doping. At suitable doping one type of spin states penetrate to the bulk of the QW and the system also enters the Quantum Anomalous Hall (QAH) state. We further confirm this through the calculations of quantum Hall conductance which shows a plateau at $e^2/h$ rather than $2e^2/h$ at such a doping state. 
		The paper predicts a certain range of controllable parameters in an inverted QW for enabling a dissipationless charge transport needed for spintronics application. 
	\end{abstract}
	
	
	\maketitle
\begin{figure}[b]
		\includegraphics[width=36.5mm,height=38.5mm]{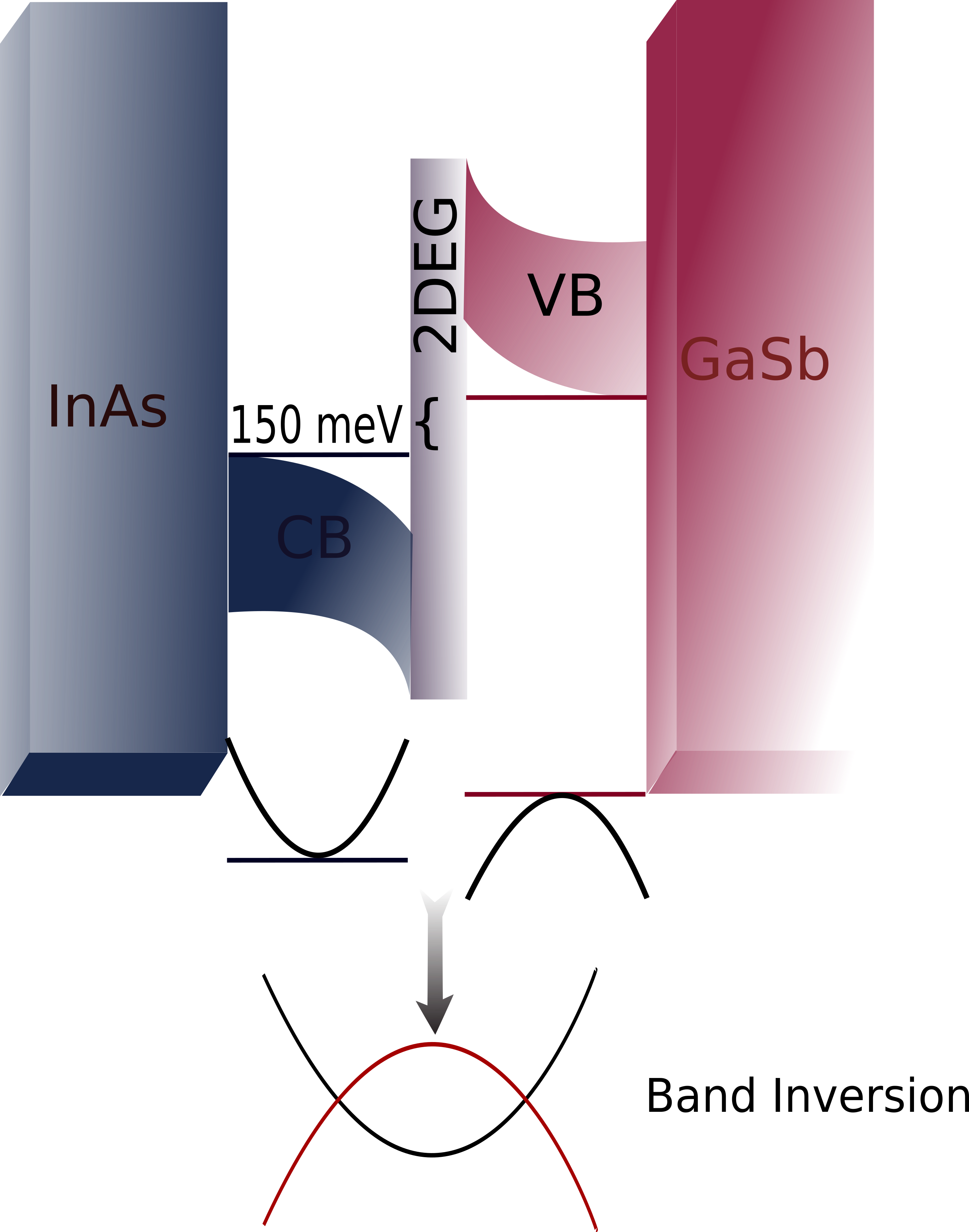}
		\includegraphics[width=36.5mm,height=36.5mm]{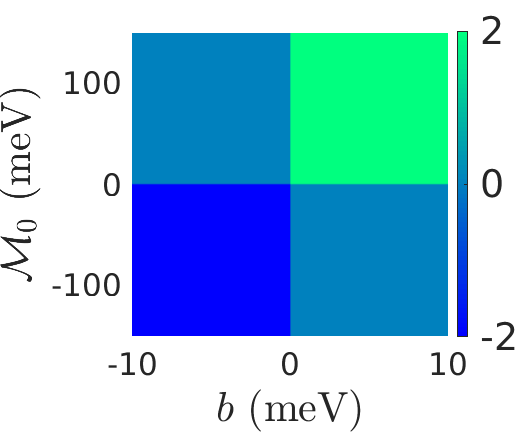}
		\caption{A schematic of the band inversion in the InAs/GaSb bilayer and the topological invariant number in such a bilayer.}
		\label{Fig1}
	\end{figure}

	\textit{\textbf{Introduction}}: 
	One of the front runners in the research of topological insulators\cite{topo,topo2,r2} is the quantum spin Hall (QSH) insulator realized in a special type of two-dimensional materials. A QSH insulator hosts counterpropagating spin-polarized edge states that are protected by time-reversal symmetry\cite{bhz,r3}.
	The gapless 2D Dirac states discovered in a band inversion 
	contact of InAs/GaSb are an exemplary material that naturally hosts these QSH states  \cite{r3,InAs1,InAs2,InAs4,ref1,ref2,ref3,ref4}. 
	In the quantum well (QW) of InAs/GaSb heterostructure, the conduction 	band of InAs is about $150$ meV lower than the valence band of GaSb providing the necessary condition to meet the band inversion\cite{r3,tune2}(Fig. [\ref{Fig1}]). Due to their robust edge states\cite{ref3,Prl1}, the QSH insulators have been at the center of condensed matter research lately. 
	The time-reversal symmetry protected pair of counter propagating edge QSH states make the QSH channel to specifically have a quantized conductance equal to $2e^2/h$, where $e^2/h\approx (25.4 \text{ k}\Omega)^{-1}$. 
	
	A magnetic field should literally destroy the QSH state forcing the two topological spin states to penetrate into the bulk of the materials\cite{TR_B}. To achieve this in an inverted InAs/GaSb QW, an in-plane magnetic field as strong as $12$ T is required. In particular, it was shown that the InAs/GaSb mesoscopic Hall channel exhibits a wide conductance plateau precisely quantized to $2e^2/h$ upto an in-plane magnetic field equals to $12$ T\cite{ref3,Prl1}. Also in the magnetic doped HgTe/CdTe QW (HgTe has been just another material that naturally has a band inversion properties), it was shown that the quantum anomalous Hall (QAH) states also can be realized.\cite{magnetic_doped} This is due to the fact that in Hg$_{1-y}$Mn$_{y}$Te, at sufficient doping, one of the spin-states will start penetrating into the bulk of the material. Note that a strong perpendicular magnetic field will also introduce the QAH phase into the InAs/GaSb heterostructure\cite{Hamil2,QAH_B1,QAH_B2,QAH_B3,Hamil3,Nat1,QAH_B4}.
	
	In this paper, we uncover these two important discoveries, from the theoretical perspective. We started with the real space tight binding Hamiltonian to describe the inverted QW. When projected it into the momentum space, a slightly modified Bernevig-Hughes-Zhang (BHZ) Hamiltonian\cite{bhz} is obtained. To take into account the magnetic Mn-doping, we added to it a Zeeman like term to the BHZ Hamiltonian. The observation in Ref. [\cite{magnetic_doped}], is confirmed through the calculation of the topological invariant number ($\mathcal{C}$) and subsequently the quantum Hall conductance. As one would expect, the number, $\mathcal{C}$, should display integers equal to $\pm 2$, $\pm 1$ and $0$ that correspond to QSH, QAH and trivial insulating phases, respectively. To address the observation of the robust QSH states in Ref. [\cite{ref3,Prl1}], we consider the effect of in-plane magnetic field added into the un-doped QW. Taking the in-plane magnetic field equal to $12$ T, as a threshold magnetic field, (after which the quantized Hall conductance $2e^2/h$ disappears),
	we predict the values of the effective in-plane Lande-\textit{g} factor in the InAs/GaSb QW, assuming it to be isotropic and field independent. 
	
	We start with the BHZ Hamiltonian which is a four-band model, having two electron bands (spin-up and spin-down) and two hole bands (spin-up and down).\cite{bhz,Hamil2,Hamil3,Hamil4,Hamil5}
	\begin{eqnarray}\label{HamilB0}
	H=\epsilon({\bf k})\sigma_0&&\tens\sigma_0+\mathcal{M}({\bf k})\sigma_0\tens\sigma_z+A k_x\sigma_z\tens\sigma_x\nonumber\\&&-A k_y\sigma_0\tens\sigma_y.
	\end{eqnarray}
	Here, $\sigma_i$, (with $i=x,y,z$) are the Pauli's spin matrices and $\sigma_0$ is the $2\times 2$ identity matrix. The band symmetry-breaking effect between that of the electron and the hole, is captured in the first term $\epsilon({\bf k})=\chi-dk^2$. The very important term $\mathcal{M}({\bf k})=\mathcal{M}_0-b k^2$ has two parameters, \textit{viz}., `$\mathcal{M}_0$' that controls the band inversion and `$b$' that symmetrically controls the band curvatures. The last two terms take care of the finite coupling between the electron and hole states. The BHZ Hamiltonian is governed by the number $\mathcal{C}=({\rm sign}[b]+{\rm sign}[\mathcal{M}_0])/2$. This number predicts the existence of the QSH states for a given choice of the set of parameters $\{\mathcal{M}_0,b\}$. 
	%

	\begin{figure}[t]
		\includegraphics[width=40.5mm,height=40.5mm]{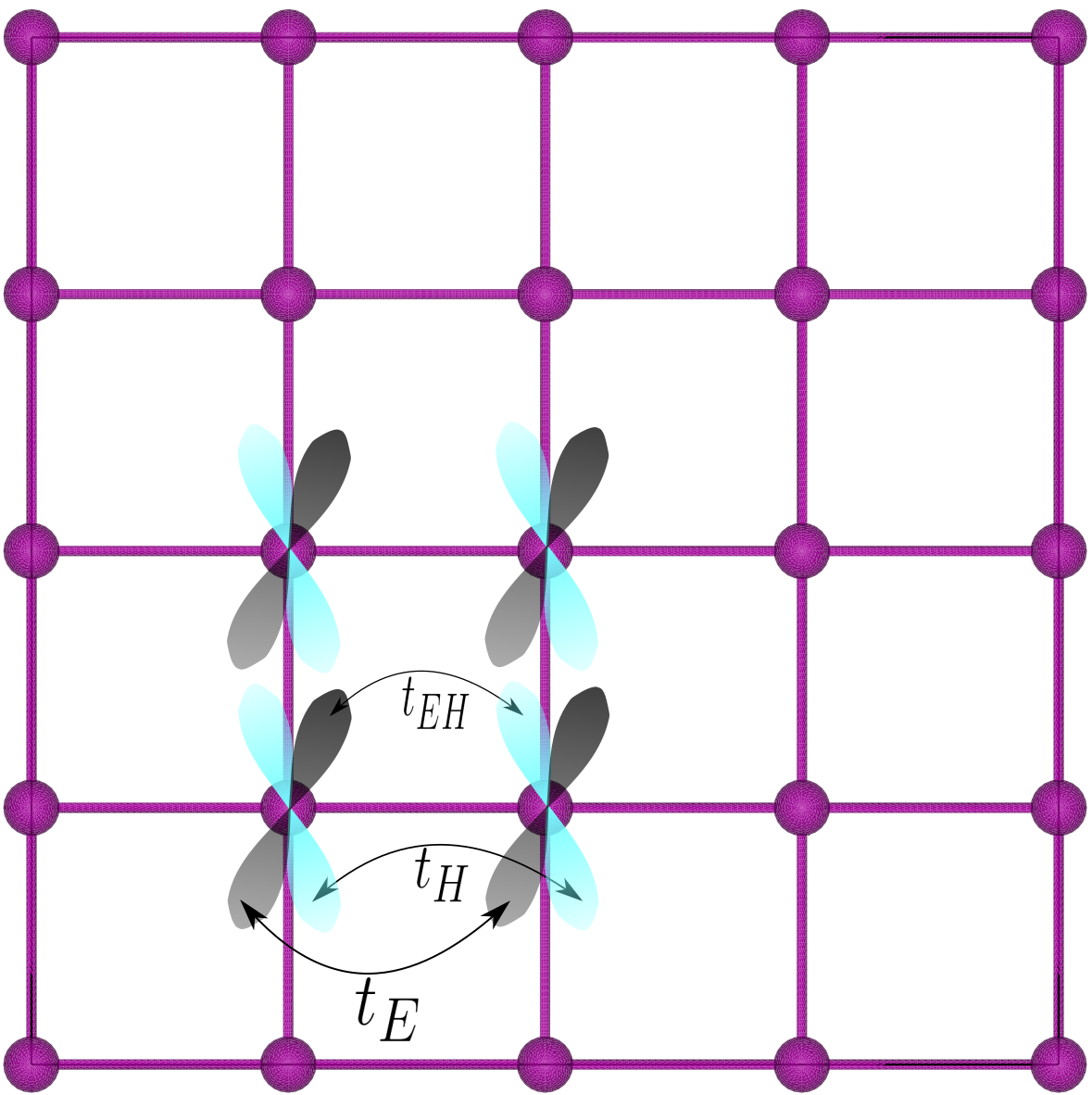}
		\caption{The intra and inter-orbitals hopping between the two orbitals ($E\; \& \; H$) projected on a  square lattice for realizing the QSH states in the InAs/GaSb QW. The $E/H$ orbital is illustrated in light black/sky-blue color.}
		\label{Fig2}
	\end{figure}

	\textit{\textbf{The lattice Model of BHZ Hamiltonian}}:
	For realization of the QSH states in such InAs/GaSb QW, we adopt the tight-binding Hamiltonian on a square lattice with two orbitals corresponding to a lattice site (Fig. [\ref{Fig2}]), represented as $E\; \& \; H$ orbitals. We can think of the two different orbitals to originate from the two III and V elements that give rise to the inverted band structure. The said tight-binding Hamiltonian is written as
	$
	H=H_0+H_{\mathcal{T}}+H_\mathcal{Z},$\cite{real_S}
	with
	\begin{equation}\label{H_real}
	\left.
	\begin{aligned}
	H_0=&\sum_{\textbf{m},\sigma}\big[\epsilon_E e^\dagger_{\textbf{m},\sigma} e_{\textbf{m},\sigma}+\epsilon_H h^\dagger_{\textbf{m},\sigma} h_{\textbf{m},\sigma}\big] \\
	H_{\mathcal{T}}=&\sum_{\langle \textbf{m},\textbf{n}\rangle,\sigma}\big[t_E e^\dagger_{\textbf{m},\sigma}e_{\textbf{n},\sigma}+t_H  h^\dagger_{\textbf{m},\sigma}h_{\textbf{n},\sigma}\big]+\\
	&\sum_{\textbf{m},\sigma}t_{EH}\big[s_z i\big( e^\dagger_{\textbf{m},\sigma}h_{\textbf{m}+\hat{x},\sigma}-e^\dagger_{\textbf{m},\sigma}h_{\textbf{m}-\hat{x},\sigma}\big)\\&+\big(e^\dagger_{\textbf{m},\sigma}h_{\textbf{m}+\hat{y},\sigma}-e^\dagger_{\textbf{m},\sigma}h_{\textbf{m}-\hat{y},\sigma}\big)\big]+h.c \\
	H_{\mathcal{Z}}&=\sum_{\textbf{m},\sigma}s_z\big[ G_Ee^\dagger_{\textbf{m},\sigma}e_{\textbf{m},\sigma}+ G_Hh^\dagger_{\textbf{m},\sigma}h_{\textbf{m},\sigma}\big]
	\end{aligned}
	\right\} 
	\end{equation}
	In the above Hamiltonian, $\epsilon_{E/H}$ denotes the onsite energy of an electron in the $E/H$ orbital. The notation $\langle {\bf m},{\bf n}\rangle$ describes a pair of the nearest neighbor (NN)
	lattice sites with position vectors ${\bf m}$ and ${\bf n}$. The terms $t_{E/H}$ and $t_{EH}$ denote the energy involed in intra-orbital and inter-orbital hopping, respectively. The magnetic doping we considered is taken into account through the term Zeeman term $H_{\mathcal{Z}}$, with $G_{E/H}$, being the resulting Zeeman energy to the $E/H$ orbital due to the doping at a lattice site. The operators $(e^\dagger_{\textbf{m},\sigma}/e_{\textbf{m},\sigma})$ \&  $(h^\dagger_{\textbf{m},\sigma}/h_{\textbf{m},\sigma})$ are the (creation/annihilation) operators for the $E$ and $H$ orbitals, respectively, with spin $\sigma$ at the site ${\bf m}$. Finally, $s_z=+/-$ for spin states $\sigma=\uparrow/\downarrow$.
	
	Writing the operator $e_{\textbf{m},\sigma}$ \&  $h_{\textbf{m},\sigma}$ in terms of the Bloch wavefunction as 
	$e_{\textbf{m},\sigma}=\sum_{\bf k}e_{\textbf{k},\sigma}\exp[i\textbf{k}\cdot \textbf{m}]$ and
	$h_{\textbf{m},\sigma}=\sum_{\bf k}h_{\textbf{k},\sigma}\exp[i\textbf{k}\cdot \textbf{m}]$, respectively, one can have
	$
	H=\sum_{{\bf k}} \psi^\dagger_{{\bf k}}H({\bf k})\psi_{{\bf k}}
	$
	where $\psi_{{\bf k}}=\begin{pmatrix}
	e_{\textbf{k},\uparrow} & h_{\textbf{k},\uparrow} & e_{\textbf{k},\downarrow} & h_{\textbf{k},\downarrow}
	\end{pmatrix}^\prime$ and $H({\bf k})$ is a slightly modified BHZ Hamiltonian as in Eq. [\ref{HamilB0}],
	with $\mathcal{M}_0\rightarrow \mathcal{M}_0+8b$ and $\chi\rightarrow \chi+8d$. The parameters $b,\;d,\;\mathcal{M}_0$, and $\chi$ in terms of $t_E,\;t_H,\;t_{EH},\;\epsilon_E,\;\epsilon_H$ are found to be
	\begin{eqnarray}
	\left.
	\begin{aligned}
	&&b=\frac{t_E-t_H}{2},\;b=\frac{t_E+t_H}{2},\;A=-2t_{EH},\\
	&&\mathcal{M}_0=\frac{\epsilon_E-\epsilon_H}{2}-2(t_E-t_H),\\
	&&\chi=\frac{\epsilon_E+\epsilon_H}{2}-2(t_E+t_H)
	\end{aligned}
	\right\} 
	\end{eqnarray}
	For HgTe, the lattice constant is $a = 0.646$ nm. For simplicity, we have taken $a=1$ $A^{\degree}$. For a typical inverted InAs/GaSb QW, the different hopping and onsite energies are $t_E=1.74$ meV, $t_H=-12$ meV, $t_{EH}=-1.9$ meV, $\epsilon_E=17$ meV and $\epsilon_H=-58$ meV.  
	Equation [\ref{HamilB0}] allows us to have a set of four energy spin-split bands given by 
	\begin{eqnarray}\label{eigen}
	\scalemath{0.91}{E_{\lambda,s_z}({\bf k})=\epsilon({\bf k})+s_z\frac{G_E+G_H}{2}+\lambda\sqrt{A^2k^2+[\mathcal{M}_{s_z}(\textbf{k})]^2}},
	\end{eqnarray}
	with their corresponding eigenstates given by
	\begin{equation}
	\left.
	\begin{aligned}
	&\psi_{{\bf k},\lambda,-1}({\bf r})=\frac{1}{\sqrt{1+{p_{\lambda,-1}({\bf k})}^2}}\begin{pmatrix}
	0\\
	0\\
	1\\
	p_{\lambda,-1}({\bf k})
	\end{pmatrix}\\
	&\psi_{{\bf k},\lambda,1}({\bf r})
	=\frac{1}{\sqrt{1+{p_{\lambda,1}({\bf k})}^2}}\begin{pmatrix}
	1\\
	p_{\lambda,1}({\bf k})\\
	0\\
	0
	\end{pmatrix}
	\end{aligned}
	\right\},
	\end{equation}
	where
	\begin{eqnarray}
	p_{\lambda,s_z}({\bf k})=\frac{-s_zAk e^{-i s_z\theta}}{\mathcal{M}_{s_z}({\bf k})+\lambda\sqrt{A^2k^2+\big[\mathcal{M}_{s_z}({\bf k})\big]^2}}.
	\end{eqnarray}
	Here, $\mathcal{M}_{s_z}({\bf k})=\mathcal{M}({\bf k})+s_z \frac{G_E-G_H}{2}$ and $\lambda=+/-$ which denotes the conduction/valence band.
	
\begin{figure}[t]
		\includegraphics[width=40.5mm,height=32.5mm]{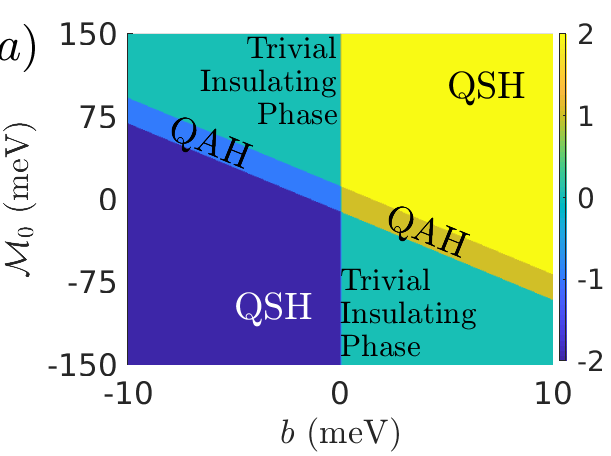}
		\includegraphics[width=42.5mm,height=30.5mm]{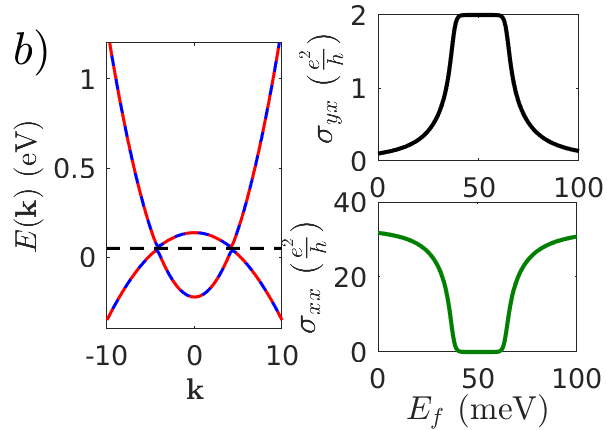}
		\includegraphics[width=42.5mm,height=30.5mm]{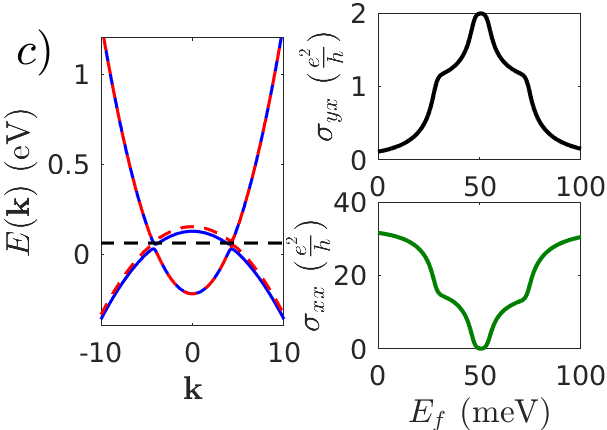}
		\includegraphics[width=42.5mm,height=30.5mm]{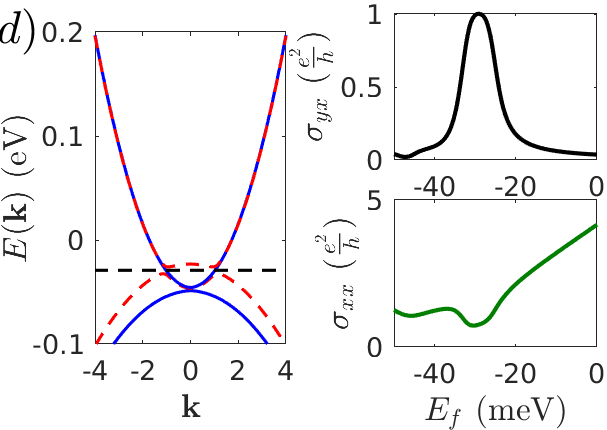}
		\includegraphics[width=72.5mm,height=6.0mm]{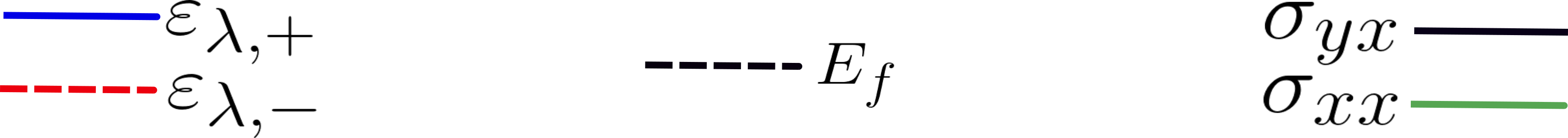}
		\caption{a). A phase diagram showing the different topological phases resulting out of the real-space tight-binding Hamiltonian in Eqn. (\ref{H_real}). We set the magnitude of the effective Zeeman energies due to doping at $G_E=12.8$ meV, $G_H=-0.35$ meV. Sections b), c) and  d) show the Hall conductance ($\sigma_{yx}$) and the longitudinal conductance ($\sigma_{xx}$) as a function of the Fermi energy with their corresponding band structure at different topological phases. For parameters in sections b) \& c) the system hosts QSH states, whereas in section d) it hosts the QAH states.}
		\label{Fig3}
	\end{figure}

	The two spin-split conduction bands ($\lambda=+$) are characterised by a Berry curvature given by
	\begin{equation}\label{berry}
	\mathcal{B}_{s_z}(\textbf{k})=i
	\frac{A^2\mathcal{M}_{s_z}({\bf k})}{2[E_{+,s_z}(\textbf{k})-E_{-,s_z}(\textbf{k})]^{3}}.
	\end{equation}
	The integral of the Berry curvature over the Brillouin zone, $\mathcal{C}_{s_z}=\int {\bf dk} \mathcal{B}_{s_z}({\bf k})/2\pi i$ gives the topological invariant number which describes the topological state of the system\cite{book}. Thus, here in the case of a magnetic doped QW, we can define two topological invariant number \textit{w.r.t.} each of the spin states
	\begin{eqnarray}
	&&\mathcal{C}_{s_z}=\frac{1}{2}\big[\text{sign} (b)+\text{sign}\{\mathcal{M}_0+8b+s_z\frac{G_E-G_H}{2}\}\big].
	\end{eqnarray}
	The topological state of the system is now defined by the number $\mathcal{C}=\mathcal{C}_{+}+\mathcal{C}_{-}$, which in the undoped case, this number could have values equal to \{-2, 0, 2\}.
	The variation of this number as a function of the parameters $\mathcal{M}_0$ and $b$ in a magnetic doped QW is shown in Fig. [\ref{Fig3} a)]. It is now easy for us to separate out three different topological phases as shown in Fig. [\ref{Fig3} a)]. The QSH phases (with $\mathcal{C}=\pm 2$) host two counterpropagating edge states with opposite spin. When the QW is magnetically doped, we saw a lifting of the Kramer degeneracy as shown in Eqn. (\ref{eigen}). With a suitable doping and for some suitable parameters ($\mathcal{M}_0 \;\& \; b$), the states ($E_{\lambda,-}$), penetrate into the bulk of the materials leaving only the states ($E_{\lambda,+}$) at the edges of the QW. In other words, the system now enters into a QAH topological phase with $\mathcal{C}_-=0$ and $\mathcal{C}_+=\pm1$. The QAH phase can also be observed when a strong magnetic field is applied perpendicular to the plane of the QW\cite{Hamil2,QAH_B1,QAH_B2,QAH_B3,Hamil3,Nat1,QAH_B4}. In this case there is no requirement of such strong magnetic field.

	\textit{\textbf{Quantum Hall conductivity and Longitudinal conductivity}}:
	To have a firm inception of the possibility of the QAH phase, we calculate the static conductivity tensor in a magnetic doped inverted QW, switching off the in-plane magnetic field. Using the Kubo formula in the linear response regime, the transversal conductivity \cite{book,ahe} in a quantum Hall channel can be written as
	\begin{eqnarray}\label{hall_f}
	&&\sigma_{yx}={\hbar}\lim_{\eta\rightarrow 0}\sum_{\lambda,s_z}\int {\bf dk}\frac{f_{\lambda,s_z}({\bf k})-f_{-\lambda,s_z}({\bf k})}{E_{\lambda,s_z}({\bf k})-E_{-\lambda,s_z}({\bf k})}\nonumber\\
	&&\times\frac{\Im \big[\langle \textbf{k},\lambda,s_z\vert \hat{J}_y\vert \textbf{k},-\lambda,s_z\rangle \langle \textbf{k},-\lambda,s_z\vert\hat{J}_x\vert \textbf{k},\lambda,s_z\rangle\big]}{E_{\lambda,s_z}({\bf k})-E_{-\lambda,s_z}({\bf k})+i\eta}.
	\end{eqnarray}
	Here, the current density operator is written as $\hat{J}_i({\bf k})=-e/\hbar\partial_{k_i}\hat{H}$ and $f_{\lambda,s_z}$ denotes the Fermi-Dirac distribution function for the quantum state $\psi_{\lambda,s_z}({\bf k})$. 
	After a fair bid of calculations, we arrive at the following expression for the static Hall conductivity
	\begin{eqnarray}\label{Hall_f1}
	\scalemath{0.93}{\sigma_{yx}=\frac{e^2}{h}\frac{1}{(2\pi)^2}\sum_{s_z}\int {\bf dk}[f_{+,s_z}({\bf k})-f_{-,s_z}({\bf k})]\mathcal{B}_{s_z}({\bf k}).}
	\end{eqnarray}
	
	To confirm the predicted resistanceless edge states, we also calculate the longitudinal conductivity whose general expression is given as follows\cite{calderin}
	\begin{eqnarray}
	&&\sigma_{xx}=\frac{\beta\hbar}{\pi}\sum_{\lambda,s_z}\int {\bf dk}\scalemath{0.94}{f_{\lambda,s_z}({\bf k})(f_{\lambda,s_z}({\bf k})-1) \delta(E_{\lambda,s_z}-E_{\lambda^\prime,s_z^\prime})}\nonumber\\
	&&\times\Re[\langle \lambda,s_z({\bf k})\vert \hat{J}_x\vert \lambda^\prime,s_z^\prime({\bf k})\rangle \langle \lambda^\prime,s_z^\prime({\bf k})\vert \hat{J}_x\vert \lambda,s_z({\bf k})\rangle].
	\end{eqnarray}
	In this case, this quantity turns out to be 
	\begin{eqnarray}\label{Long}
	&&\sigma_{xx}=\frac{e^2}{\pi\hbar}\frac{\beta}{\Gamma}\sum_{\lambda,s_z}\int kdk f_{\lambda,s_z}(f_{\lambda,s_z}-1)\nonumber\\&&\times k^2\scalemath{0.94}{\frac{(A^2-2b \mathcal{M}_{s_z}({\bf k})-\lambda 2 d\sqrt{A^2k^2+[\mathcal{M}_{sz}({\bf k})]^2})^2}{A^2k^2+[\mathcal{M}_{sz}({\bf k})]^2}}.
	\end{eqnarray}

	\textit{\textbf{Results and Discussions}}:
	For analysing the results of our calculations, we take the parameters as shown in the table below
	\begin{table}[H]
		\centering
		\begin{tabular}{|c | c| c|c | c| c|}
			\hline
			\rowcolor{lightgray}
			\thead{$b$\\ (meV a$^2$)}  & 	\thead{$d$ \\(meV a$^2$)}& \thead{$\chi$ \\(eV)}& \thead{$A$ \\(meV a)}& \thead{$\mathcal{M}_0$ \\(meV)}\\
			\hline
			$\thead{686}$  & $\thead{-512}$ & $0$ & \thead{$36.5$}   & \thead{$10$}\\ \hline
		\end{tabular}
		\caption{The different parameters associated with the BHZ Hamiltonian in eq. (\ref{HamilB0})}
		\label{Table_1}
	\end{table}
	
	We now analyse the characteristics of $\sigma_{yx}$ and $\sigma_{xx}$ obtained in Eqns. [\ref{Hall_f1} \& \ref{Long}]. The variation of these quantities as a function of the Fermi energy is shown in Fig. [\ref{Fig3} (b-d)]. For choosing a specific set of parameters so as to restrict ourselves in a certain topological region, we refer to the phase diagram in Fig. [\ref{Fig3} a)]. The sections [$b),\;\& \; c)$] correspond to a specific set of parameters from the QSH region and thus one can witness the Hall conductivity plateaues at $2\frac{e^2}{h}$ as a consequence of the existing two counterpropagating edge states. Since the system is subjected to a magnetic doping, there is a possibility that the system will also enter the QAH states with one of two spin states that remains topological (refer to Fig. [\ref{Fig3} (d)]). When the system approaches the QAH topological phases, the states $E_{\lambda,-}$ penetrate into the bulk of the material, [their band structure shown in blue color of Fig. [\ref{Fig3}] d)], whereas the states $E_{\lambda,+}$ remains topological with the band inversion traits still persisting. However, the QAH states appear over a thin regions in the phase diagram as shown in Fig. [\ref{Fig3} a)].  Also, there exists an unwanted region where all the states will creep into the bulk and the system behaves like a trivial insulator, with $\mathcal{C}=0$. Here both the states $E_{\lambda,-}$ and $E_{\lambda,+}$ lose all their band inversion traits.

	\begin{figure}[t]
		\includegraphics[width=40.5mm,height=38.5mm]{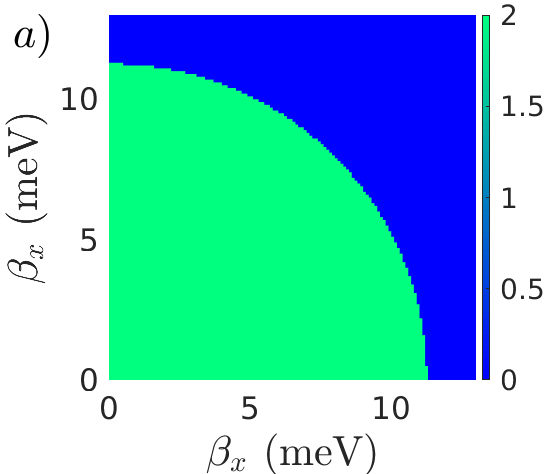}
		\includegraphics[width=40.5mm,height=38.5mm]{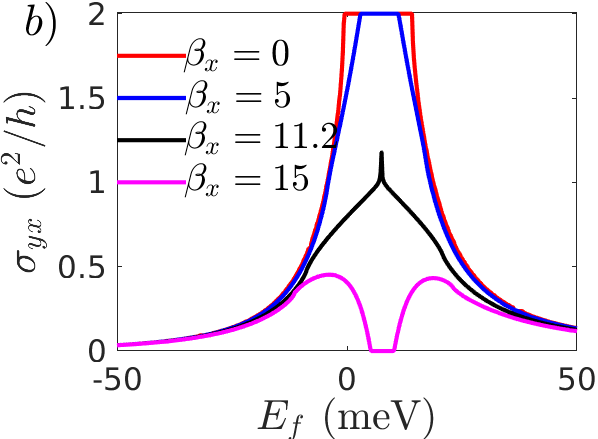}
		\caption{a) A figure showing the topological invariant number as a function of the in-plane magnetic field. b) The plot showing the variation of the hall conductance as a function of the Fermi enery of a set of values of $\beta_x$.}
		\label{Fig4}
	\end{figure} 
	
	To complement the results obtained in the Ref. [\cite{ref3}], we switch on the in-plane magnetic field in undoped HgTe qunatum well, ${\bf B}=B_x \hat{x}+B_y\hat{y}$. We now have a Zeeman term $H_{\rm in-plane}=g^\ast (\sigma_x \beta_x + \sigma_y \beta_y)$, (assuming the Lande-\textit{g} factor to be isotropic and field-independent) added to Eqn. [\ref{HamilB0}]. Here $\beta_{x,y}=\mu_B B_{x,y}$, with $\mu_B$ being the Bohr magneton. Taking this into account the Berry curvature now is transformed to
	\begin{eqnarray}
	&&\mathcal{B}({\bf k})=i \frac{A[A(\mathcal{M}_0+b\textbf{k}^2)+2b(k_x\beta_x-k_y\beta_y)]}{2[E_{+}(\textbf{k})-E_{-}(\textbf{k})]^{3}}.
	\end{eqnarray}
	Note that the energy dispersion here becomes anisotropic in the momentum space and that the Kramers degeneracy is not lifted. Similar calculations of the Hall and longitudinal conductances as in Eqns. [\ref{Hall_f1} \& \ref{Long}] can also be obtained in this case.
	
	We now carried out the numerical analysis for the Hall conductivity when the in-plane magnetic field is switch on. We choose the same set of system parameters from Table. [\ref{Table_1}]. Note that, unlike a perpendicular magnetic field, the in-plane magnetic field does not introduce the QAH phase into the system. Our point of interest here is the threshold magnetic field beyond which the quantized Hall plateau ($2e^2/h$) diminishes. We first plotted the phase diagram in Fig. [\ref{Fig4} a)] to help us identifying the threshold magnetic field. 
	In the reference \cite{ref3}, this threshold magnetic field was found to be $12 $ T. For simplicity, we consider a case where the magetic field is applied only along the x-axis. A careful observation of Fig. [\ref{Fig4} a)], indicates that this threshold field occured for $\beta_{x}^{\rm threshold}=11.2$ meV. This results in an effective in-plane Lande-\textit{g}$_x$ factor of $16.1$. To accertain the fact that the Hall plateau disappeared for any in-plane magnetic field whose magnitude is greater than $\beta_{x}^{\rm threshold}$, we plot the Hall conductance in Fig. [\ref{Fig4} b)]. We choose a set of four values of $\beta_{x}=\{0 , 5, 11.2 ,15\}$ meV, two of which $\beta_{x}=\{0,5\}$ meV are well within the green region of Fig. [\ref{Fig4} a)] indicating a quantized Hall conductance of $2 e^2/h$. Analysing of the Hall conductance \textit{w.r.t.} this set of values of $\beta_{x}$ shows a complete disappearance of Hall plateau after a threshold value of $\beta_{x}> \beta_{x}^{\rm threshold}=11.2$ meV. 

	In \textit{\textbf{conclusions}}, we have studied the different possible topological phases in an inverted InAs/GaSb QW. We consider two cases for our study. Firstly, when the QW is magnetically doped and secondly, when we turned on the external in-plane magnetic field on an undoped QW. The calculations in this paper are based on a tight binding Hamiltonian projected on a square lattice. A given lattice site is considered to have the contribution from the two different orbitals of the two constituent atoms that provide the inverted band structure. In a suitable doped QW, the resulting phase diagram predicted two kinds of topological phases, namely, the QAH phase and the QSH phase. Expectedly, our results show two possible quantized Hall conductance of $e^2/h$ (for QAH phase) and $2e^2/h$ (for QSH phase). The QAH phase occurred due to the fact that one of the spin states starts penetrating into the bulk of the QW at the suitable doping condition and its band inversion properties are lost. In the second part of this paper, we have presented a precise calculation of the Lande-\textit{g} factor in such a QW with no doping considered. Our calculation is based on the recent discoveries that the QSH states exist in an applied in-plane magnetic field as strong as 12 T. The paper predicts a certain range of controllable parameters in an inverted QW for enabling a dissipationless charge transport needed for spintronics application.

	\textit{Acknowledgments}: This work is an outcome of
	the Research work carried out under the DST-INSPIRE project DST/INSPIRE/04/2019/000642, Government of India.


\begin{thebibliography}{55}
		
		\bibitem{topo}
		C. L. Kane and E. J. Mele, \href{https://journals.aps.org/prl/abstract/10.1103/PhysRevLett.95.226801}{Phys. Rev. Lett. \textbf{95}, 226801 (2005).}
		
		\bibitem{topo2}
		L. Fu, C. L. Kane, and E. J. Mele,
		\href{https://journals.aps.org/prl/abstract/10.1103/PhysRevLett.98.106803}{Phys. Rev. Lett, \textbf{98}, 106803 (2007).}
		
		\bibitem{r2}
		C. L. Kane and E. J. Mele, \href{https://journals.aps.org/prl/abstract/10.1103/PhysRevLett.95.146802}{Phys. Rev. Lett. \textbf{95}, 146802
		(2005).}
		
		\bibitem{bhz}
		\href{https://www.science.org/doi/10.1126/science.1133734}{B.A. Bernevig, T.L. Hughes, and S.-C. Zhang, 	Science \textbf{314}, 1757 (2006).}
		
		\bibitem{r3}
		B. A. Bernevig and S.-C. Zhang, \href{https://journals.aps.org/prl/abstract/10.1103/PhysRevLett.96.106802}{Phys. Rev. Lett. \textbf{96},
		106802 (2006).}
		
		
		\bibitem{InAs1}
		I. Knez, R. R. Du, and G. Sullivan, \href{https://journals.aps.org/prl/abstract/10.1103/PhysRevLett.107.136603}{Phys. Rev. Lett. \textbf{107}, 136603
		(2011).}
		
		\bibitem{InAs2}
		I. Knez, R. R. Du, and G. Sullivan, \href{https://journals.aps.org/prl/abstract/10.1103/PhysRevLett.109.186603}{Phys. Rev. Lett. \textbf{109}, 186603
		(2012).}
		
		
		\bibitem{InAs4}
		Dong-Hui Xu, Jin-Hua Gao, Chao-Xing Liu, Jin-Hua Sun, Fu-Chun Zhang, and Yi Zhou
		\href{https://journals.aps.org/prb/pdf/10.1103/PhysRevB.89.195104}{Phys. Rev. B \textbf{89}, 195104 (2014).}
		
		\bibitem{ref1}
		C. Liu, T. L. Hughes, X.-L. Qi, K. Wang, and S.-C. Zhang, \href{https://journals.aps.org/prl/abstract/10.1103/PhysRevLett.100.236601}{Phys. Rev. Lett. \textbf{100}, 236601 (2008).}
		
		\bibitem{ref2}
		K. Suzuki, Y. Harada, K. Onomitsu, and K. Muraki, \href{https://journals.aps.org/prb/abstract/10.1103/PhysRevB.87.235311}{Phys. Rev. B. \textbf{87}, 235311 (2013).}
		
		\bibitem{ref3}
		L. Du, I. Knez, G. Sullivan, and R.-R. Du, \href{https://journals.aps.org/prl/abstract/10.1103/PhysRevLett.114.096802}{Phys. Rev. Lett. \textbf{114}, 096802 (2015).}
		
		\bibitem{ref4}
		V. S. Pribiag, A. J. A. Beukman, F. Qu, M. C. Cassidy, C. Charpentier, W. Wegscheider, and L. P. Kouwenhoven, \href{https://www.nature.com/articles/nnano.2015.86}{Nat. Nanotechnol. \textbf{10}, 593-597 (2015).}
		
		
%
		
		\bibitem{tune2}
		J. G. Checkelsky, Y. S. Hor, R. J. Cava, and N. P. Ong
		\href{https://journals.aps.org/prl/abstract/10.1103/PhysRevLett.106.196801}{Phys. Rev. Lett. \textbf{106}, 196801 (2011).}
		
		
		\bibitem{Prl1}
		E. M. Spanton, K. C. Nowack, L. Du, G. Sullivan, R.-R. Du, and K. A. Moler
		\href{https://journals.aps.org/prl/abstract/10.1103/PhysRevLett.113.026804}{Phys. Rev. Lett. \textbf{113}, 026804 (2014).}
		
		\bibitem{TR_B}
		 Kubo, R., \href{https://journals.jps.jp/doi/abs/10.1143/JPSJ.12.570}{J. Phys. Soc. Jpn.
		\textbf{12}, 570-586 (1957)}
		
		\bibitem{magnetic_doped}
		C.-X. Liu, X.-L. Qi, X. Dai, Z. Fang, and S.-C. Zhang, \href{https://journals.aps.org/prl/abstract/10.1103/PhysRevLett.101.146802}{Phys. Rev. Lett. \textbf{101}, 146802 (2008).}
		
		\bibitem{QAH_B1}
		S. S. Krishtopenko, S. Ruffenach, F. Gonzalez-Posada, G. Boissier, M. Marcinkiewicz, M. A. Fadeev, A. M. Kadykov, V. V. Rumyantsev, S. V. Morozov, V. I. Gavrilenko, C. Consejo, W. Desrat,
		B. Jouault, W. Knap, E. Tournie, and F. Teppe1,
		\href{https://journals.aps.org/prb/abstract/10.1103/PhysRevB.97.245419}{Phys. Rev. B \textbf{97}, 245419 (2018).}
		
		\bibitem{QAH_B2}
		A. Zakharova, S. T. Yen, and K. A. Chao,
		\href{https://journals.aps.org/prb/abstract/10.1103/PhysRevB.69.115319}{Phys. Rev. B \textbf{69}, 115319 (2004).}
		
		\bibitem{QAH_B3}
		A. Mawrie, \href{https://iopscience.iop.org/article/10.1088/1361-648X/ac5fd7}{J. Phys. Condens. Matter. \textbf{34}, 245301 (2022).}
		
		
		\bibitem{Nat1}
		E. Y. Ma, M. R. Calvo, J. Wang, B. Lian, M. Muhlbauer, C. Brune, Y.-T. Cui, K. Lai, W. Kundhikanjana, Y. Yang, M. Baenninger, M. Konig, C. Ames, H. Buhmann, P. Leubner, L. W. Molenkamp, S.-C. Zhang, D. G. -Gordon, M. A. Kelly, and Z.-X. Shen,
		\href{https://www.nature.com/articles/ncomms8252}{Nature Communications \textbf{6}, 7252 (2015).}
		
		
		\bibitem{Hamil3}
		M. Konig, H. Buhmann, L. W. Molenkamp, T. Hughes, C. X. Liu, X. L. Qi, and S.-C. Zhang, \href{https://journals.aps.org/prb/pdf/10.1103/PhysRevB.82.184516}{J. Phys. Soc. Jpn. \textbf{77}, 031007 (2008).}
		
		
		\bibitem{Hamil2}
		M. Konig, S. Wiedmann, C. Brune, A. Roth, H. Buhmann, L. W. Molenkamp, X. L. Qi, and S.-C. Zhang, \href{https://www.science.org/doi/10.1126/science.1148047}{Science 318, 766 (2007).}
		
		\bibitem{QAH_B4}
		J.-c. Chen, J. Wang, and Q.-f. Sun, \href{https://journals.aps.org/prb/abstract/10.1103/PhysRevB.85.125401}{Phys. Rev. B \textbf{85}, 125401 (2012).}
		
		
		
		
		
		
		\bibitem{Hamil4}
		B. Scharf, A. M. Abiague, and J. Fabian, \href{https://journals.aps.org/prb/abstract/10.1103/PhysRevB.86.075418}{Phys. Rev. B, \textbf{86}, 075418 (2012).}
		
		
		\bibitem{Hamil5}
		W. Beugeling, C. X. Liu, E. G. Novik, L. W. Molenkamp,
		and C. Morais Smith, \href{https://journals.aps.org/prb/abstract/10.1103/PhysRevB.85.195304}{Phys. Rev. B, \textbf{85}, 195304 (2012).}
		
		\bibitem{real_S}	
		Z. Chen and T. K. Ng, 
		Phys. Rev. B, \href{https://journals.aps.org/prb/abstract/10.1103/PhysRevB.85.195304}{Phys. Rev. B, \textbf{99}, 235157 (2019).}
		
		
		\bibitem{book}
		S.-Q. Shen, \href{https://link.springer.com/book/10.1007/978-3-642-32858-9}{Topological Insulators (Springer-Verlag,
		Heidelberg, 2013)}
		
		
		
		\bibitem{ahe}
		K. Imura, A. Yamakage, S.J. Mao, A. Hotta, Y. Kuramoto, \href{https://journals.aps.org/prb/abstract/10.1103/PhysRevB.82.085118}{Phys. Rev. \textbf{82}, 085118 (2010).}
		
		
		\bibitem{calderin}
		L. Calderin, V. V. Karasiev, and S. B. Trickey, \href{https://www.sciencedirect.com/science/article/pii/S0010465517302539}{Comput.
		Phys. Commun. \textbf{221}, 118 (2017).}
		%
		
		
		%
		%
		%
		%
		%
		%
		%
		%
		
		
		
		
		
		
	\end{thebibliography}
\end{document}